\documentclass[12pt]{iopart}
\usepackage{iopams}  
\usepackage{bbold}
\usepackage{color}
\usepackage{graphicx}

\newcommand{\la}{\langle}
\newcommand{\ra}{\rangle}
\newcommand{\lla}{\left\langle}
\newcommand{\rra}{\right\rangle}

\newcommand{\mcP}{\mathcal{P}}
\newcommand{\mcC}{\mathcal{C}}
\newcommand{\One}{\mathbb{1}}

\newcommand{\tsum}{\textstyle\sum}     
\newcommand{\tprod}{\textstyle\prod}

\begin{document}

\title[Projection probabilities for random states] {Joint probability 
  distributions for projection probabilities of random orthonormal states}

\author{L. Alonso and T. Gorin}
\address{Departamento de F\'\i sica, Universidad de Guadalajara, Guadalajara, 
  Jalisco, M\' exico}
\ead{lazarus.alon@gmail.com}
\vspace{10pt}
\begin{indented}
\item[] September 2015
\end{indented}

\begin{abstract}
A finite dimensional quantum system for which the quantum chaos conjecture 
applies has eigenstates, which show the same statistical properties than the 
column vectors of random orthogonal or unitary matrices. Here, we consider the 
different probabilities for obtaining a specific outcome in a projective 
measurement, provided the system is in one of its eigenstates. We then give
analytic expressions for the joint probability density for these 
probabilities, with respect to the ensemble of random matrices. In the case of 
the unitary group, our results can be applied, also, to the phenomenon of 
universal conductance fluctuations, where the same mathematical quantities 
describe partial conductances in a two-terminal mesoscopic scattering problem 
with a finite number of modes in each terminal.
\end{abstract}

%
%
%

\section{\label{I} Introduction}

Let $\mathcal{H}$ be the Hilbert space of finite dimension 
$N={\rm dim}(\mathcal{H})$ corresponding to some quantum system. Let $S$ be 
a subspace in $\mathcal{H}$, with $K={\rm dim}(S)$. We choose an orthonormal
basis $\mathcal{B}= \{ \, \varphi_j\, \}_{1\le j\le N}$ in $\mathcal{H}$, with 
the first $K$ elements lying in $S$. Then, the projection of a normalized state 
$\psi\in\mathcal{H}$ on the subspace $S$ has the squared norm 
\begin{equation}
t= \sum_{j=1}^K |\la\varphi_j|\psi\ra|^2 \; .
\label{I:normsquaredproj}\end{equation}
In a projective measurement, the observable to be measured is associated to a 
Hermitian operator $\hat A$, which always has an orthonormal basis of 
eigenstates. Assume $\mathcal{B}$ is such a basis and $S$ is the eigenspace 
corresponding to a $K$-fold degenerate eigenvalue of $\hat A$, then $t$ as 
given in (\ref{I:normsquaredproj}) is the probability that the outcome of 
the measurement of $\hat A$ is that eigenvalue.

In this paper, we consider several orthonormal states $\psi_1, \ldots \psi_R$ 
and the squared norms of their projections
\begin{equation}
t_\xi = \sum_{j=1}^K |\la\varphi_j|\psi_\xi\ra|^2 \; , \qquad 
1\le \xi \le R \; .
\label{I:txiprojs}\end{equation}
We calculate the joined probability distribution of these quantities, provided 
the states $\psi_1, \ldots \psi_R$ a chosen at random from one of two invariant 
ensembles. In order to define these ensembles, we use the basis $\mathcal{B}$, 
which allows to write the quantum states as complex column vectors of length 
$N$, and the orthogonal (unitary) group as real (complex) unitary 
$N$$\times$$N$ matrices~\cite{Weyl39}. Then, the first ensemble is defined as 
the set of collections of $R$ orthonormal column vectors equipped with the 
unique probability measure, which is invariant under orthogonal 
transformations.
Similarly, the second ensemble is defined as the set of collections of $R$ 
orthonormal column vectors equipped with the unique probability measure, which 
is invariant under unitary transformations.

Both, the orthogonal and the unitary group may be turned into an ensemble of
random matrices, using the normalized Haar measure as probability 
measure~\cite{Haar33,Meh2004}. Unless states otherwise, we refer to these 
ensembles simply by their group names, $O(N)$ and $U(N)$. It then turns out 
that the two ensembles defined above can be obtained from $O(N)$ and $U(N)$,
by selecting only $R$ column vectors (we may always choose the first $R$ column 
vectors) from a given group element. Below, we will no longer distinguish
between the original ensembles of collections of $R$ column vectors and the 
ensembles defined over the whole $N$$\times$$N$ matrices. 

The groups $O(N)$ and $U(N)$ as invariant ensembles have become an important
reference in the study of broad classes of quantum systems, first in nuclear
physics~\cite{Bro81}, later in quantum chaos~\cite{Haake2001,Stoe99} and 
mesoscopic quantum transport~\cite{Bee97,MelKum04}, and more recently in 
quantum information; e.g.~\cite{GS02b,GS03,PinSel15}. To give ye another 
example, a general definition of an entanglement measure for mixed states is 
based on the so called ``convex roof'' construction, where mutually orthogonal 
vectors are used to parameterize the different possibilities to write the 
density matrix in question as a mixture of pure states~\cite{Min05,Hor09}. Our 
results may thus be useful in any of these areas.

Averages over matrix elements of the orthogonal and the unitary group have
been considered in different contexts. Averages over monomials are considered
in~\cite{Aub03,ColSni06,GL08}. The statistical properties of eigenvalues of
$M$$\times$$M$ sub-matrices ($M < N$) in~\cite{ZycSom00,Nov07}. Ensembles of 
scattering matrices may be derived from the unitary group, 
also~\cite{MelSel80}. In this area, there are connections to the distribution 
of transmission eigenvalues~\cite{MePeKu88}, and the probability distributions 
of individual scattering matrix elements in the so called Heidelberg
model~\cite{Kum13,NoKSG14}.

The paper is organized as follows. In the next section, we introduce our 
notation and define the quantities of interest. There, we also treat the 
one-vector case, which has been solved much earlier. In section~\ref{VR} we 
derive the general expressions for an arbitrary number of column vectors.
In section~\ref{V2} we treat the case $R=2$ in more detail. In that case, one
can often evaluate all the integrals, to arrive at very simple expressions 
which contain only rational and/or algebraic functions. In this section, we 
also discuss two physical applications, the distribution of probabilities for
measurement outcomes in a closed quantum system, and partial conductances in 
systems showing universal conductance fluctuations. Conclusions are provided 
in section~\ref{C}.

\section{\label{D} Definitions and notation}

We will denote the joint probability densities of the variables 
$t_1,\ldots,t_R$ defined in (\ref{I:txiprojs}) as $P_{NK}(t_1,\ldots,t_R)$
in the case the average is over the orthogonal group and as 
$\mcP_{NK}(t_1,\ldots,t_R)$ in the case of the unitary group. It is 
understood that the probability density is with respect to the flat measure 
$\rmd t_1\, \ldots \rmd t_R$ in the real unit hyper-cube of dimension $R$. 
With the help of the Dirac delta function, we write the following formal 
expressions for these probability densities
\begin{eqnarray}
P_{NK}(t_1,\ldots,t_R) = \lla {\tprod_{\xi=1}^R}\, 
   \delta\Big (t_\xi - {\tsum_{j=1}^K w_{j\xi}^2}\Big )\rra_{O(N)} \; ,
\label{D:PNKON}\\
\mcP_{NK}(t_1,\ldots,t_R) = \lla {\tprod_{\xi=1}^R}\, 
   \delta\Big (t_\xi - {\tsum_{j=1}^K |w_{j\xi}|^2}\Big )\rra_{U(N)}\; .
\label{D:PNKUN}\end{eqnarray}
Here, $\bi{w}$ denotes a group element, i.e. a $N$$\times$$N$ matrix with real
or complex entries. The matrix elements are denoted by $w_{j\xi}$, where we
use Latin (Greek) indices for rows (columns). The angular brackets denote
the ensemble average over the respective group, with respect to the respective
normalized Haar measure~\cite{Haar33}. We may write these ensemble averages as
integrals over the flat space of all matrix elements, implementing the
orthonormality conditions on the column vectors by additional delta functions.
Originally, this idea is due to Ullah~\cite{UllPor63,Ullah64} and more recently
it had been used to calculate group averages of monomials~\cite{Gor02,GL08}.

\subsection{\label{V1} One point functions}

In the case of the one point functions, we consider the probability density
for the projection of only one random vector on the $K$-dimensional subspace.
The result is known for a long time, see e.g.~\cite{Bro81}. Nevertheless,
we present our calculation in some detail, since it is different from
the usual approach and since it introduces some techniques used again for the 
general case.

\subsubsection*{Orthogonal case}

For the orthogonal case, the desired probability density may be written as 
\begin{eqnarray}
P_{NK}(t) &= \lla 
   \delta\Big (t - {\tsum_{j=1}^K w_j^2}\Big )\rra_{O(N)} 
 = C_N\int\rmd\Omega(\bi{w})\; 
    \delta\Big (t - {\tsum_{j=1}^K w_j^2}\Big ) \nonumber\\
&= C_N\int\rmd^N\bi{w}\; \delta\big (1 - \|\bi{w}\|^2)\;
    \delta\Big (t - {\tsum_{j=1}^K w_j^2}\Big ) \; .
\end{eqnarray}
In this equation, $\rmd\Omega(\bi{w})$ is the invariant measure on the unit 
hyper-sphere in $\mathbb{R}^N$, $\rmd^N\bi{w}$ is the flat measure in 
$\mathbb{R}^N$, and $\|\bi{w}\|^2 = \sum_{j=1}^N w_j^2$. The normalization 
constant (to be determined below) is denoted by $C_N$, as it depends on $N$ 
but not on $K$. We follow \cite{Ullah64} to eliminate the delta-function which 
implements the normalization, and apply the transformation 
$w_j\to u_j= \sqrt{r}\, w_j$ for an arbitrary parameter $r>0$. This gives
\begin{equation}
P_{NK}(t)\; r^{N/2-2} = C_N\int\rmd^N\bi{u}\; 
    \delta(r - \|\bi{u}\|^2)\;
    \delta\Big (rt - {\tsum_{j=1}^K u_j^2}\Big ) \; .
\end{equation}
Multiplying both sides with $\rme^{-r}$ and integrating $r$ from $0$ to 
$\infty$ then yields
\begin{equation}
P_{NK}(t)\; \Gamma(N/2-1) = C_N\int\rmd^N\bi{u}\; 
    \delta\Big (\|\bi{u}\|^2\, t - {\tsum_{j=1}^K u_j^2}\Big )\; 
    \rme^{-\|\bi{u}\|^2}\; .
\end{equation}
The normalization constant, $C_N$, is obtained from the requirement that 
\begin{equation}
1 = \int_0^1\rmd t\; P_{NK}(t) = C_N\int\rmd^N\bi{w}\; 
    \delta(1 - \|\bi{w}\|^2) \; .
\end{equation}
Applying the same trick as above, allows to evaluate the integral right away.
It results in $C_N= \Gamma(N/2)/\pi^{N/2}$. Returning to $P_{NK}(t)$, we
replace the last delta function by its Fourier representation. 
\begin{eqnarray}
P_{NK}(t) &= \frac{N/2 -1}{\pi^{N/2}}
   \int\frac{\rmd s}{2\pi}\int\rmd^N\bi{u}\; \rme^{-\|\bi{u}\|^2 (1+\rmi st)}\; 
   \prod_{j=1}^K \rme^{\rmi s\, u_j^2} \nonumber\\
&= \frac{N/2 -1}{\pi^{N/2}} \int\frac{\rmd s}{2\pi}\int\rmd^N\bi{u}\; 
   \prod_{j=K+1}^N \rme^{-u_j^2 (1+\rmi st)}\; \prod_{j=1}^K
   \rme^{-u_j^2 [1-\rmi s(1-t)]} \nonumber\\
&= \frac{N-2}{4\pi}\int\frac{\rmd s}{\sqrt{(1+\rmi st)^{N-K}\;
      [1-\rmi s\, (1-t)]^K}} \; .
\label{V1:Oresult}\end{eqnarray}
This integral can be evaluated as explained in the appendix, with the result 
given in (\ref{app:Imk}). Therefore,
\begin{equation}
P_{NK}(t) = I_{(N-K)/2,K/2}(t,1-t) = 
   \frac{\Gamma(N/2)\; t^{K/2 -1}\; (1-t)^{(N-K)/2 -1}}
        {\Gamma((N-K)/2)\, \Gamma(K/2)}\; .
\label{V1:Oresult2}\end{equation}
It is easy to show that this result is in agreement with similar results on
the statistics of random vector components reviewed in \cite{Bro81}.

\subsubsection*{Unitary case}

For the unitary case, the probability density for the squared norm of the 
projection on a $K$-dimensional subspace may be written as 
\begin{eqnarray}
\mcP_{NK}(t) &= \lla 
   \delta\Big (t - {\tsum_{j=1}^K |w_j|^2}\Big )\rra_{U(N)} \nonumber\\
 &= \mcC_N\int\rmd^{2N}\bi{w}\; \delta\big (1 - \|\bi{w}\|^2)\;
    \delta\Big (t - {\tsum_{j=1}^K |w_j|^2}\Big ) \; .
\end{eqnarray}
The integral is now over the $2N$ dimensional space of real and imaginary
parts of the complex components of the vector $\bi{w}$. The normalization
restricts the integration to the unit hyper-sphere in this space. The sum in 
the second delta function, goes equally over the squares of real and imaginary 
parts of the vector $\bi{w}$. This simply means that 
$\mcP_{NK}(t)= P_{2N,2K}(t)$ such that 
\begin{equation}
\mcP_{NK}(t)= \frac{N-1}{2\pi}\int
   \frac{\rmd s}{(1+\rmi st)^{N-K}\, (1-\rmi s\, (1-t))^K}\; .
\label{V1:Uresult}\end{equation}
In this case, we may again use (\ref{app:Imk}) to find
\begin{equation}
\mcP_{NK}(t)= I_{N-K,K}(t,1-t) = \frac{\Gamma(N)\; t^{K-1}\; (1-t)^{N-K-1}}
                                    {\Gamma(N-K)\; \Gamma(K)} \; .
\label{V1:Uresult2}\end{equation}

\section{\label{VR} General \boldmath $R$ point functions}

In this section, we derive general integral expressions for the 
case of an arbitrary number of vectors. Again, we start with the orthogonal
case and treat the unitary case afterwards. Let us adopt a few conventions 
which simplify the interpretation of the following expressions which often
involve multiple integrals: (i) The symbol for integration together with the 
expression which denotes the integration measure form one unit, and the 
integand then extends up to the next plus or minus sign. (ii) The symbol for
multiple products, acts on the expression to its right, extending up to the 
next plus or minus sign. To restrict the symbol's scope otherwise, we use 
brackets surrounding the product term and the product symbol. The curly 
brackets below, in (\ref{VRO:PNKdef}), are used in that way.

\subsection{\label{VRO} Orthogonal case}

For the joint probability density as defined in (\ref{D:PNKON}) we write
\begin{equation}
\fl P_{NK}(t_1,\ldots,t_R) = C_{NR}\left\{ {\tprod_{\xi=1}^R} 
   \int\rmd\Omega(\bi{w}_\xi)\; \delta\Big ( t_\xi - {\tsum_{j=1}^K} w_{j\xi}^2
   \Big )\right\} \; \prod_{\mu < \nu}^R 
   \delta\big ( \la\bi{w}_\mu\, |\, \bi{w}_\nu\ra \big ) \; ,
\label{VRO:PNKdef}\end{equation}
where $\rmd\Omega(\bi{w}_\xi)$ denotes the uniform measure on the hyper-sphere 
in $\mathbb{R}^N$. The last product of delta functions implements the 
orthogonality condition between the column vectors $\bi{w}_\xi$ of the elements 
of the orthogonal group. 

\subsubsection*{Normalization} 
Before treating the full expression, let us calculate the normalization 
constant.
\begin{eqnarray}
\fl C_{NR}^{-1} = \left\{ {\tprod_{\xi=1}^R} 
   \int\rmd^N\bi{w}_\xi\; \delta( 1 - \|\bi{w}_\xi\|^2)\right\}\; 
   \prod_{\mu < \nu}^R \delta\big ( \la\bi{w}_\mu\, |\, \bi{w}_\nu\ra \big ) 
\nonumber\\
 = \left\{ {\tprod_{\xi=1}^R} 
   \int\rmd^N\bi{u}_\xi\; r_\xi^{-N/2 +1 + (R-1)/2}\; 
   \delta( r_\xi - \|\bi{u}_\xi\|^2)\right\}\; 
   \prod_{\mu < \nu}^R \delta\big ( \la\bi{u}_\mu\, |\, \bi{u}_\nu\ra \big )\; ,
\end{eqnarray}
from which it follows that
\begin{eqnarray}
\fl C_{NR}^{-1}\; \Gamma[(N-R+1)/2]^R = \left\{ {\tprod_{\xi=1}^R} 
   \int\rmd^N\bi{u}_\xi\; \rme^{-\|\bi{u}_\xi\|^2}\right\}\; 
   \prod_{\mu < \nu}^R \delta\big ( \la\bi{u}_\mu\, |\, \bi{u}_\nu\ra \big )
\nonumber\\
 = \left\{ {\tprod_{\mu < \nu}^R}\int\frac{\rmd\tau_{\mu\nu}}{2\pi}\right\}\;
   \left\{ {\tprod_{\xi=1}^R} 
   \int\rmd^N\bi{u}_\xi\; \rme^{-\|\bi{u}_\xi\|^2}\right\}\;
   \prod_{\mu < \nu}^R 
   \rme^{-\rmi\tau_{\mu\nu}\, \la\bi{u}_\mu\, |\, \bi{u}_\nu\ra} \nonumber\\
 = \left\{ {\tprod_{\mu < \nu}^R}\int\frac{\rmd\tau_{\mu\nu}}{2\pi}\right\}\;
   {\tprod_{j=1}^N}\int\rmd^R\bi{u}_j'\; 
   \rme^{- \la\bi{u}_j'|\bi{D}\, \bi{u}_j'\ra}\; .
\label{VR:ONnorm1}\end{eqnarray}
In this expression and below we denote with 
$\bi{u}'_j= (u_{j1},\ldots,u_{jR})^T$ the $j$'th row vector of the orthogonal
matrix $\bi{u}$, restricted to the first $R$ components. Correspondingly, we 
denote the scalar product between two row vectors $\bi{u}'$ and $\bi{v}'$ as 
$\la\bi{u}'|\bi{v}'\ra$. Hence, with
\begin{equation}
\bi{D} = \One + \frac{\rmi}{2}\left(\begin{array}{ccc} 
                                   0 & \tau_{12} & \ldots \\
                                   \tau_{12} & 0 & \\
                             \vdots & & \ddots\end{array}\right)
\end{equation}
we find that 
\begin{equation}
\la\bi{u}_j'|\bi{D}\, \bi{u}_j'\ra 
 = \sum_{\mu\nu} u_{j\mu}\, D_{\mu\nu}\, u_{j\nu} 
 = \|\bi{u}_j'\|^2 + \rmi \sum_{\mu < \nu} \tau_{\mu\nu}\, u_{j\mu} u_{j\nu}\; .
\end{equation}
Here, $\|\bi{u}_j'\|^2 = \la\bi{u}_j'|\bi{u}_j'\ra$ denotes the squared norm
of the row vector $\bi{u}_j'$. The integrals over the $\bi{u}_j'$ in 
(\ref{VR:ONnorm1}) are standard Gaussian integrals, which can be evaluated 
in terms of the determinant of $\bi{D}$. In this way, we obtain
\begin{equation}
C_{NR}^{-1}\; \Gamma[(N-R+1)/2]^R = \left\{ {\tprod_{\mu < \nu}^R}
   \int\frac{\rmd\tau_{\mu\nu}}{2\pi}\right\}\; 
   \frac{\pi^{NR/2}}{\det(\bi{D})^{N/2}} \; .
\label{VRO:CNRnormconst}\end{equation}
  
\subsubsection*{Full expression}
Returning to the full expression (\ref{VRO:PNKdef}), we start again by removing 
the delta functions implementing the normalization of the column vectors. From
\begin{eqnarray}
\fl P_{NK}(t_1,\ldots,t_R) = C_{NR}\left\{ {\tprod_{\xi=1}^R} 
   \int\rmd^N\bi{w}_\xi\; \delta(1- \|\bi{w}_\xi\|^2)\; 
   \delta\Big ( t_\xi - {\tsum_{j=1}^K} w_{j\xi}^2
   \Big )\right\} \; \prod_{\mu < \nu}^R 
   \delta\big ( \la\bi{w}_\mu\, |\, \bi{w}_\nu\ra \big ) \nonumber\\
 = C_{NR}\left\{ {\tprod_{\xi=1}^R} \int\rmd^N\bi{u}_\xi\; 
   r^{-N/2 +2 +(R-1)/2}\; \delta(r_\xi - \|\bi{u}_\xi\|^2)\; 
   \delta\Big ( r_\xi t_\xi - {\tsum_{j=1}^K} u_{j\xi}^2 \Big )\right\} 
\nonumber\\
 \qquad\times \prod_{\mu < \nu}^R \delta\big ( \la\bi{u}_\mu\, |\, \bi{u}_\nu\ra 
 \big ) \; ,
\end{eqnarray}
it follows that
\begin{eqnarray}
\fl P_{NK}(t_1,\ldots,t_R)\; \Gamma[(N-R-1)/2]^R = C_{NR}
  \left\{ {\tprod_{\mu < \nu}^R}\int\frac{\rmd\tau_{\mu\nu}}{2\pi}\right\}
\nonumber\\
  \qquad\times
  \left\{ {\tprod_{\xi=1}^R} \int\frac{\rmd s_\xi}{2\pi}\int\rmd^N\bi{u}_\xi\; 
    \rme^{-\|\bi{u}_\xi\|^2 (1+\rmi s_\xi t_\xi)}\; 
    \rme^{\rmi s_\xi \sum_{j=1}^K u_{j\xi}^2} \right\} \; \prod_{\mu < \nu}^R 
   \rme^{-\rmi\tau_{\mu\nu}\, \la\bi{u}_\mu\, |\, \bi{u}_\nu\ra} \nonumber\\
= C_{NR}
  \left\{ {\tprod_{\mu < \nu}^R}\int\frac{\rmd\tau_{\mu\nu}}{2\pi}\right\}
  \left\{ {\tprod_{\xi=1}^R} \int\frac{\rmd s_\xi}{2\pi}\right\}
\nonumber\\
  \qquad\times\left\{ {\tprod_{j=1}^K}\int\rmd^R\bi{u}_j'\; 
     \rme^{-\sum_{\xi=1}^R u_{j\xi}^2 [1-\rmi s_\xi (1-t_\xi)] 
           -\rmi\sum_{\mu <\nu}^R u_{j\mu}\, \tau_{\mu\nu}\, u_{j\nu}} 
\right\}\nonumber\\
 \qquad\times\left\{ {\tprod_{j=K+1}^N}\int\rmd^R\bi{u}_j'\; 
     \rme^{-\sum_{\xi=1}^R u_{j\xi}^2 (1+\rmi s_\xi t_\xi) 
           -\rmi\sum_{\mu <\nu}^R u_{j\mu}\, \tau_{\mu\nu}\, u_{j\nu}} 
\right\} \; .
\end{eqnarray}
Then, with the help of the matrices 
\begin{equation}
\fl \bi{A} = \One + \rmi\left(\begin{array}{ccc} 
                         s_1 t_1 & \tau_{12}/2 & \ldots \\
                         \tau_{12}/2 & s_2 t_2 & \\
                       \vdots & & \ddots\end{array}\right)\; , \quad
\bi{B} = \One + \rmi\left(\begin{array}{ccc} 
                     -s_1 (1-t_1) & \tau_{12}/2 & \ldots \\
                     \tau_{12}/2 & -s_2 (1-t_2) & \\
                   \vdots & & \ddots\end{array}\right)\; ,
\end{equation}
we may write that
\begin{eqnarray}
\fl P_{NK}(t_1,\ldots,t_R)\; \Gamma[(N-R-1)/2]^R = C_{NR}
  \left\{ {\tprod_{\mu < \nu}^R}\int\frac{\rmd\tau_{\mu\nu}}{2\pi}\right\}
  \left\{ {\tprod_{\xi=1}^R} \int\frac{\rmd s_\xi}{2\pi}\right\}
\nonumber\\
 \qquad\times\left\{ {\tprod_{j=K+1}^N}\int\rmd^R\bi{u}_j'\; 
     \rme^{- \la\bi{u}_j'|\bi{A}\, \bi{u}_j'\ra} \right\}
 \left\{ {\tprod_{j=1}^K}\int\rmd^R\bi{u}_j'\; 
     \rme^{- \la\bi{u}_j'|\bi{B}\, \bi{u}_j'\ra} \right\} \nonumber\\
= C_{NR}\left\{ {\tprod_{\xi=1}^R} \int\frac{\rmd s_\xi}{2\pi}\right\}
  \left\{ {\tprod_{\mu < \nu}^R}\int\frac{\rmd\tau_{\mu\nu}}{2\pi}\right\}
  \frac{\pi^{NR/2}}{\det(\bi{A})^{(N-K)/2}\, \det(\bi{B})^{K/2}} \; .
\end{eqnarray}
In view of the result for the normalization constant $C_{NR}$ in 
(\ref{VRO:CNRnormconst}) and the fact that at $\bi{s}= \bi{0}$, it holds that 
$\bi{A}= \bi{B}= \bi{D}$. Thus, we define
\begin{equation}
Z(\bi{s})= 
  \left\{ {\tprod_{\mu < \nu}^R}\int\frac{\rmd\tau_{\mu\nu}}{2\pi}\right\}
  \frac{1}{\det(\bi{A})^{(N-K)/2}\, \det(\bi{B})^{K/2}} \; ,
\label{VRO:Zgen}\end{equation}
which allows to write
\begin{equation}
P_{NK}(t_1,\ldots,t_R) = \frac{[(N-R-1)/2]^R}{Z(\bi{0})} 
   \left\{ {\tprod_{\xi=1}^R} \int\frac{\rmd s_\xi}{2\pi}\right\}\; 
   Z(\bi{s}) \; .
\label{VRO:genform}\end{equation}

\subsubsection*{One vector case}
Strictly speaking, (\ref{VRO:genform}) applies for $R > 1$, only.  However, for 
$R=1$, $\bi{A} = 1 +\rmi st$ and $\bi{B}= 1 -\rmi s (1-t)$, such that 
(\ref{VRO:Zgen}) may be interpreted as
\begin{equation}
Z(s)= \frac{1}{(1 +\rmi\, st)^{(N-K)/2}\, [1 -\rmi s (1-t)]^{K/2}} \; , \qquad
Z(0)= 1\; .
\end{equation}
This yields
\begin{equation}
P_{NK}(t) = \frac{N-2}{2}\int\frac{\rmd s}{2\pi}\; 
   \frac{1}{(1 +\rmi\, st)^{(N-K)/2}\, [1 -\rmi s (1-t)]^{K/2}}\; ,
\end{equation}
in agreement with (\ref{V1:Oresult}).

\subsection{\label{VRU} Unitary case}

For the joint probability density as defined in (\ref{D:PNKUN}) we write
\begin{equation}
\fl \mcP_{NK}(t_1,\ldots,t_R) = \mcC_{NR}\left\{ {\tprod_{\xi=1}^R} 
   \int\rmd\Omega_2(\bi{w}_\xi)\; 
   \delta\Big ( t_\xi - {\tsum_{j=1}^K} |w_{j\xi}|^2 \Big )\right\}\; 
   \prod_{\mu < \nu}^R 
   \delta^2\big ( \la\bi{w}_\mu\, |\, \bi{w}_\nu\ra \big ) \; .
\label{VRU:PNKdef}\end{equation}
Here, $\rmd\Omega_2(\bi{w}_\xi)$ denotes the 
uniform measure on the hyper-sphere in $\mathbb{R}^{2N}$. The last product of
delta functions implements the orthogonality conditions between the column 
vectors $\bi{w}_\xi$ of the elements of the unitary group. These are 
two-dimensional because for $\la\bi{w}_\mu | \bi{w}_\nu\ra$ to be zero, the 
real and the imaginary part must be zero.

\subsubsection*{Normalization} 

Before treating the full expression, let us calculate the normalization 
constant.
\begin{eqnarray}
\fl \mcC_{NR}^{-1} = \left\{ {\tprod_{\xi=1}^R} 
   \int\rmd^{2N}\bi{w}_\xi\; \delta( 1 - \|\bi{w}_\xi\|^2)\right\}\; 
   \prod_{\mu < \nu}^R \delta^2( \la\bi{w}_\mu\, |\, \bi{w}_\nu\ra ) 
\nonumber\\
 = \left\{ {\tprod_{\xi=1}^R} 
   \int\rmd^{2N}\bi{u}_\xi\; r_\xi^{-N +1 + R-1}\; 
   \delta( r_\xi - \|\bi{u}_\xi\|^2)\right\}\; 
   \prod_{\mu < \nu}^R \delta^2( \la\bi{u}_\mu\, |\, \bi{u}_\nu\ra )\; ,
\end{eqnarray}
from which it follows
\begin{eqnarray}
\fl \mcC_{NR}^{-1}\; \Gamma(N-R+1)^R = \left\{ {\tprod_{\xi=1}^R} 
   \int\rmd^{2N}\bi{u}_\xi\; \rme^{-\|\bi{u}_\xi\|^2}\right\}\; 
   \prod_{\mu < \nu}^R \delta^2( \la\bi{u}_\mu\, |\, \bi{u}_\nu\ra )
\nonumber\\
 = \left\{ {\tprod_{\mu < \nu}^R}\int\frac{\rmd^2\tau_{\mu\nu}}{4\pi^2}\right\}
   \; \left\{ {\tprod_{\xi=1}^R} 
   \int\rmd^{2N}\bi{u}_\xi\; \rme^{-\|\bi{u}_\xi\|^2}\right\}\;
   \prod_{\mu < \nu}^R 
   \rme^{-\rmi\, {\rm Im}(\tau_{\mu\nu}\, \la\bi{u}_\mu\, |\, \bi{u}_\nu\ra)}\; , 
\end{eqnarray}
where we have used the Fourier representation of the two-dimensional delta
function with complex argument, defined in (\ref{app:2Dcomplexdelta}).
As in the orthogonal case, it will again prove convenient to change from
column vectors with Greek indices to row vectors with Latin indices. The
notations for vectors, scalar products and the vector norm are analogous to the
orthogonal case. However, the coefficients are now complex such that 
$\la\bi{u}_j'|\bi{u}_k'\ra = \sum_{\xi=1}^R u_{j\xi}^*\, u_{k\xi}$. With this, 
we may write
\begin{eqnarray}
\fl \sum_{\xi=1}^R \|\bi{u}_\xi\|^2 +\rmi \sum_{\mu < \nu}^R 
   {\rm Im}(\tau_{\mu\nu}\, \la\bi{u}_\mu\, |\, \bi{u}_\nu\ra)
 = \sum_{j=1}^N \Big [\, \|\bi{u}_j'\|^2 + \frac{1}{2} {\tsum_{\mu < \nu}^R}
  (\tau_{\mu\nu}\, u_{j\mu}^* u_{j\nu} - \tau_{\mu\nu}^*\, u_{j\mu} u_{j\nu}^*)
  \, \Big ] \nonumber\\
\fl\qquad\qquad
 = \sum_{j=1}^N \Big [\, \|\bi{u}_j'\|^2 + \frac{1}{2} {\tsum_{\mu < \nu}^R}
   u_{j\mu}^*\, \tau_{\mu\nu}\, u_{j\nu} - \frac{1}{2} {\tsum_{\nu < \mu}^R} 
   u_{j\mu}^*\, \tau_{\nu\mu}^*\, u_{j\nu} \, \Big ]
 = \la\bi{u}_j'|\, \bi{G}\, \bi{u}_j'\ra \; , 
\end{eqnarray}
where
\begin{equation}
\bi{G} = \One + \frac{1}{2}\left(\begin{array}{ccc} 
                                   0 & \tau_{12} & \ldots \\
                                   -\tau_{12}^* & 0 & \\
                             \vdots & & \ddots\end{array}\right)
 = \One + \frac{\rmi}{2}\left(\begin{array}{ccc} 
                                   0 & -\rmi\tau_{12} & \ldots \\
                                    \rmi\tau_{12}^* & 0 & \\
                             \vdots & & \ddots\end{array}\right)\; .
\end{equation}
We thus find
\begin{eqnarray}
 \mcC_{NR}^{-1}\; \Gamma(N-R+1)^R &=
 \left\{ {\tprod_{\mu < \nu}^R}\int\frac{\rmd\tau_{\mu\nu}}{4\pi^2}\right\}\;
   {\tprod_{j=1}^N}\int\rmd^{2R}\bi{u}_j'\; 
   \rme^{- \la\bi{u}_j'|\bi{G}\, \bi{u}_j'\ra}\nonumber\\
 &= \left\{ {\tprod_{\mu < \nu}^R}\int\frac{\rmd\tau_{\mu\nu}}{4\pi^2}\right\}\;
    \frac{\pi^{NR}}{\det(\bi{G})^N} \; . 
\label{VR:UNnorm}\end{eqnarray}

\subsubsection*{Full expression}
Returning to the full expression (\ref{VRU:PNKdef}), we again remove first the
delta functions implementing the normalization of the column vectors. Hence,
\begin{eqnarray}
\fl \mcP_{NK}(t_1,\ldots,t_R) = \mcC_{NR}\left\{ {\tprod_{\xi=1}^R} 
   \int\rmd^{2N}\bi{w}_\xi\; \delta(1- \|\bi{w}_\xi\|^2)\; 
   \delta\Big ( t_\xi - {\tsum_{j=1}^K} |w_{j\xi}|^2 \Big )\right\}\; 
   \prod_{\mu < \nu}^R 
   \delta^2( \la\bi{w}_\mu\, |\, \bi{w}_\nu\ra ) \nonumber\\
 = \mcC_{NR}\left\{ {\tprod_{\xi=1}^R} \int\rmd^{2N}\bi{u}_\xi\; 
   r^{-N +2 +(R-1)}\; \delta(r_\xi - \|\bi{u}_\xi\|^2)\; 
   \delta\Big ( r_\xi t_\xi - {\tsum_{j=1}^K} |u_{j\xi}|^2 \Big )\right\} 
\nonumber\\
 \qquad\times\prod_{\mu < \nu}^R \delta^2(\la\bi{u}_\mu\, |\, \bi{u}_\nu\ra )\; ,
\end{eqnarray}
from which it follows that
\begin{eqnarray}
\fl \mcP_{NK}(t_1,\ldots,t_R)\; \Gamma(N-R)^R = \mcC_{NR}
  \left\{ {\tprod_{\mu < \nu}^R}\int\frac{\rmd^2\tau_{\mu\nu}}{4\pi^2}\right\}
\nonumber\\
  \qquad\times \left\{ {\tprod_{\xi=1}^R} 
      \int\frac{\rmd s_\xi}{2\pi}\int\rmd^{2N}\bi{u}_\xi\; 
      \rme^{-\|\bi{u}_\xi\|^2 (1+\rmi s_\xi t_\xi)}\; 
    \rme^{\rmi s_\xi \sum_{j=1}^K |u_{j\xi}|^2} \right\} \; 
    \prod_{\mu < \nu}^R 
    \rme^{-\rmi\, {\rm Im}(\tau_{\mu\nu}\, \la\bi{u}_\mu\, |\, \bi{u}_\nu\ra)} 
\nonumber\\
= \mcC_{NR}
  \left\{ {\tprod_{\mu < \nu}^R}\int\frac{\rmd^2\tau_{\mu\nu}}{4\pi^2}\right\}
  \left\{ {\tprod_{\xi=1}^R} \int\frac{\rmd s_\xi}{2\pi}\right\}
\nonumber\\
  \qquad\times\left\{ {\tprod_{j=1}^K}\int\rmd^{2R}\bi{u}_j'\; 
     \rme^{-\sum_{\xi=1}^R |u_{j\xi}|^2 [1-\rmi s_\xi (1-t_\xi)] 
            -\rmi/2\sum_{\mu <\nu}^R (u_{j\mu}^*\, \tau_{\mu\nu}\, u_{j\nu}
               - u_{j\mu}\, \tau_{\mu\nu}^*\, u_{j\nu}^*)} \right\}\nonumber\\
 \qquad\times\left\{ {\tprod_{j=K+1}^N}\int\rmd^{2R}\bi{u}_j'\; 
     \rme^{-\sum_{\xi=1}^R |u_{j\xi}|^2 (1+\rmi s_\xi t_\xi) 
            -\rmi/2\sum_{\mu <\nu}^R (u_{j\mu}^*\, \tau_{\mu\nu}\, u_{j\nu}
               - u_{j\mu}\, \tau_{\mu\nu}^*\, u_{j\nu}^*)} \right\} \; .
\end{eqnarray}
Again, we may define matrices 
\begin{equation}
\fl \bi{E} = \One + \rmi\left(\begin{array}{ccc} 
                         s_1 t_1 & -\rmi\tau_{12}/2 & \ldots \\
                         \rmi\tau_{12}^*/2 & s_2 t_2 & \\
                       \vdots & & \ddots\end{array}\right)\; , \quad
\bi{F} = \One + \rmi\left(\begin{array}{ccc} 
                     -s_1 (1-t_1) & -\rmi\tau_{12}/2 & \ldots \\
                     \rmi\tau_{12}^*/2 & -s_2 (1-t_2) & \\
                   \vdots & & \ddots\end{array}\right)\; ,
\end{equation}
such that 
\begin{eqnarray}
\fl \mcP_{NK}(t_1,\ldots,t_R)\; \Gamma(N-R)^R = \mcC_{NR}
  \left\{ {\tprod_{\mu < \nu}^R}\int\frac{\rmd^2\tau_{\mu\nu}}{4\pi^2}\right\}
  \left\{ {\tprod_{\xi=1}^R} \int\frac{\rmd s_\xi}{2\pi}\right\}
\nonumber\\
 \qquad\times\left\{ {\tprod_{j=K+1}^N}\int\rmd^{2R}\bi{u}_j'\; 
     \rme^{- \la\bi{u}_j'|\bi{E}\, \bi{u}_j'\ra} \right\}
 \left\{ {\tprod_{j=1}^K}\int\rmd^{2R}\bi{u}_j'\; 
     \rme^{- \la\bi{u}_j'|\bi{F}\, \bi{u}_j'\ra} \right\} \nonumber\\
= \mcC_{NR}\left\{ {\tprod_{\xi=1}^R} \int\frac{\rmd s_\xi}{2\pi}\right\}
  \left\{ {\tprod_{\mu < \nu}^R}\int\frac{\rmd^2\tau_{\mu\nu}}{4\pi^2}\right\}
  \frac{\pi^{NR}}{\det(\bi{E})^{N-K}\, \det(\bi{F})^K} \; .
\end{eqnarray}
In view of the result for the normalization constant $\mcC_{NR}$ in 
(\ref{VR:UNnorm}), and because $\bi{E}= \bi{F}= \bi{G}$, at $\bi{s}= \bi{0}$, 
we define
\begin{equation}
Z(\bi{s})= 
  \left\{ {\tprod_{\mu < \nu}^R}\int\frac{\rmd^2\tau_{\mu\nu}}{4\pi^2}\right\}
  \frac{1}{\det(\bi{E})^{N-K}\, \det(\bi{F})^K} \; ,
\end{equation}
which allows to write
\begin{equation}
\mcP_{NK}(t_1,\ldots,t_R) = \frac{(N-R)^R}{Z(\bi{0})} 
   \left\{ {\tprod_{\xi=1}^R} \int\frac{\rmd s_\xi}{2\pi}\right\}\; 
   Z(\bi{s}) \; .
\label{VRU:genform}\end{equation}

\subsubsection*{One vector case}
In a similar manner as for the orthogonal group, we may write for $R=1$,
\begin{equation}
Z(s)= \frac{1}{(1 +\rmi\, st)^{N-K}\, [1 -\rmi s (1-t)]^K} \; , \qquad
Z(0)= 1\; ,
\end{equation}
such that 
\begin{equation}
\mcP_{NK}(t) = (N-1)\int\frac{\rmd s}{2\pi}\; 
   \frac{1}{(1 +\rmi\, st)^{N-K}\, [1 -\rmi s (1-t)]^K}\; ,
\end{equation}
in agreement with (\ref{V1:Uresult}).

\section{\label{V2} Two-point functions, examples and applications}

In this section, we consider the case of $R=2$ column vectores, and thus the 
joint probability density $P_{NK}(t_1,t_2)$. That is the probability density
of two probabilitites $t_1$ and $t_2$, where $t_1$ ($t_2$) is the probability 
for finding the system in a given $K$ dimensional subspace when it is 
prepared in one (another) eigenstate. In practice, each realization of the 
system leads to one unitary matrix, representing the eigenstates, and each 
choice of two eigenstates leads to a pair of probabilities $(t_1,t_2)$. 
Averaging over many different systems then yields the joint probability density
$P_{NK}(t_1,t_2)$.

Again, we will treat the orthogonal group and the unitary group in separate
subsections. In both cases it will be convenient to use the abbreviations 
$\alpha_j = 1+\rmi s_j t_j$ and $\beta_j= 1-\rmi s_j (1-t_j)$.

\subsection{\label{V2O} Orthogonal case}

Starting again from the general expression (\ref{VRO:genform}), we find for 
$R=2$ that
\begin{eqnarray}
\det(\bi{A})= (1+\rmi s_1 t_1)\, (1+\rmi s_2 t_2) + \tau^2/4
 = \alpha_1\, \alpha_2 + \tau^2/4\; , \nonumber\\
\det(\bi{B})= [1-\rmi s_1 (1-t_1)]\, [1-\rmi s_2 (1-t_2)] + \tau^2/4
 = \beta_1\, \beta_2 + \tau^2/4\; ,
\end{eqnarray}
and
\begin{equation}
Z(\bi{s})= \int\frac{\rmd\tau}{2\pi}\; \frac{1}{
 (\alpha_1\, \alpha_2 + \tau^2/4)^{(N-K)/2}\, 
 (\beta_1\, \beta_2 + \tau^2/4)^{K/2}} \; .
\end{equation}
Here, it turns out to be more convenient to postpone the integration over 
$\tau$, and evaluate the integral over $s_2$ first.
\begin{eqnarray}
\fl \int\frac{\rmd s_2}{2\pi}\; Z(\bi{s})= \int\frac{\rmd\tau}{2\pi}\;
  \frac{1}{\alpha_1^{(N-K)/2}\, \beta_1^{K/2}}\int\frac{\rmd s_2}{2\pi}\; 
  \frac{1}{(\alpha_1' +\rmi s_2 t_2)^{(N-K)/2}\,
     [\beta_1' -\rmi s_2 (1-t_2) ]^{K/2}}\nonumber\\
= \frac{2}{N-2}\; \frac{1}{\alpha_1^{(N-K)/2}\, \beta_1^{K/2}}\;
  \int\frac{\rmd\tau}{2\pi}\; \frac{I_{(N-K)/2,K/2}(t_2,1-t_2)}
    {[\beta_1' t_2 + \alpha_1' (1-t_2)]^{N/2-1}}\nonumber\\
= \frac{2\, P_{NK}(t_2)}{N-2}\; \frac{1}{\alpha_1^{(N-K)/2}\, \beta_1^{K/2}}\;
  \int\frac{\rmd\tau}{2\pi}\;
  \frac{1}{[\beta_1' t_2 + \alpha_1' (1-t_2)]^{N/2-1}} \; ,
\end{eqnarray}
where $\alpha_1' = 1 + \tau^2/(4\alpha_1)$, $\beta_1'= 1 + \tau^2/(4\beta_1)$, 
and where we have used (\ref{app:Iabmk}) and (\ref{app:PNKImk}). We then find
that
\begin{eqnarray}
\fl \int\frac{\rmd s_2}{2\pi}\; Z(\bi{s})= \frac{2\, P_{NK}(t_2)}{N-2}\;
   \frac{1}{\alpha_1^{(N-K)/2}\, \beta_1^{K/2}}\;
  \int\frac{\rmd\tau}{2\pi}\;
  \frac{1}{[1 + \tau^2\, (t_2/\beta_1 + (1-t_2)/\alpha_1)/4]^{N/2-1}}
\nonumber\\
= \frac{2\, P_{NK}(t_2)}{N-2}\;
  \frac{1}{\alpha_1^{(N-K)/2}\, \beta_1^{K/2}}\; \frac{1}{\sqrt{\pi}}\; 
 \frac{\Gamma[(N-3)/2]}{\sqrt{t_2/\beta_1 + (1-t_2)/\alpha_1}\; \Gamma(N/2-1)}
\nonumber\\
= \frac{\Gamma[(N-3)/2]\; P_{NK}(t_2)}{\sqrt{\pi}\, \Gamma(N/2)}\; 
   \frac{1}{\alpha_1^{(N-K-1)/2}\, \beta_1^{(K-1)/2}}\;
   \frac{1}{\sqrt{t_2\, \alpha_1 + (1-t_2)\, \beta_1}} \; ,
\end{eqnarray}
where we used the integration formula (\ref{app:LorentzInt}) for real 
Lorentzian integrals. From this it follows 
\begin{eqnarray}
P_{NK}(t_1,t_2)= 
   \frac{\Gamma[(N-3)/2]\; P_{NK}(t_2)}{\sqrt{\pi}\, \Gamma(N/2)}\;
   \frac{(N-3)^2/4}{Z(\bi{0})} \nonumber\\
 \qquad\times\int\frac{\rmd s_1}{2\pi}\;
   \frac{1}{\alpha_1^{(N-K-1)/2}\, \beta_1^{(K-1)/2} 
            \sqrt{1 +\rmi s_1 (t_1+t_2 -1)}} \; .
\end{eqnarray}
With
\begin{equation}
Z(\bi{0})= \int\frac{\rmd\tau}{2\pi} \frac{1}{(1+\tau^2/4)^{N/2}} 
 = \frac{\Gamma[(N-1)/2]}{\sqrt{\pi}\, \Gamma(N/2)} \; ,
\end{equation}
we therefore obtain
\begin{equation}
\fl\qquad\quad P_{NK}(t_1,t_2)= P_{NK}(t_2)\; \frac{N-3}{4\pi}\int
   \frac{\rmd s_1}{\alpha_1^{(N-K-1)/2}\, \beta_1^{(K-1)/2} 
            \sqrt{1 +\rmi s_1 (t_1+t_2 -1)}} \; .
\label{V2O:finres}\end{equation}

\subsection{\label{V2U} Unitary case}

For the unitary group, we start from the general expression (\ref{VRU:genform}) 
and for $R=2$, we find
\begin{eqnarray}
\det(\bi{E})= (1+\rmi s_1 t_1)\, (1+\rmi s_2 t_2) + |\tau|^2/4
 = \alpha_1\, \alpha_2 + |\tau|^2/4\; , \nonumber\\
\det(\bi{F})= [1-\rmi s_1 (1-t_1)]\, [1-\rmi s_2 (1-t_2)] + |\tau|^2/4
 = \beta_1\, \beta_2 + |\tau|^2/4\; ,
\end{eqnarray}
and therefore
\begin{equation}
Z(\bi{s})= \int\frac{\rmd^2\tau}{4\pi^2}\; \frac{1}{
 (\alpha_1\, \alpha_2 + |\tau|^2/4)^{N-K}\, 
 (\beta_1\, \beta_2 + |\tau|^2/4)^K} \; .
\end{equation}
Again, we will evaluate the integral over $s_2$ before that over $\tau$. Thus,
\begin{eqnarray}
\fl \int\frac{\rmd s_2}{2\pi}\; Z(\bi{s})= \int\frac{\rmd^2\tau}{4\pi^2}\;
  \frac{1}{\alpha_1^{N-K}\, \beta_1^K}\int\frac{\rmd s_2}{2\pi}\; 
  \frac{1}{(\alpha_1' +\rmi s_2 t_2)^{N-K}\,
     [\beta_1' -\rmi s_2 (1-t_2) ]^K}\nonumber\\
= \frac{1}{N-1}\; \frac{1}{\alpha_1^{N-K}\, \beta_1^K}\;
  \int\frac{\rmd^2\tau}{4\pi^2}\; \frac{I_{N-K,K}(t_2,1-t_2)}
    {[\beta_1' t_2 + \alpha_1' (1-t_2)]^{N-1}}\nonumber\\
= \frac{\mcP_{NK}(t_2)}{N-1}\; \frac{1}{\alpha_1^{N-K}\, \beta_1^K}\;
  \int\frac{\rmd^2\tau}{4\pi^2}\;
  \frac{1}{[\beta_1' t_2 + \alpha_1' (1-t_2)]^{N-1}} \; ,
\end{eqnarray}
where $\alpha_1' = 1 + |\tau|^2/(4\alpha_1)$,
$\beta_1'= 1 + |\tau|^2/(4\beta_1)$, and where we have used (\ref{app:Iabmk}) 
and (\ref{app:PNKImk}) once more. Hence,
\begin{eqnarray}
\fl \int\frac{\rmd s_2}{2\pi}\; Z(\bi{s})= \frac{\mcP_{NK}(t_2)}{N-1}\;
   \frac{1}{\alpha_1^{N-K}\, \beta_1^K}\; \int\frac{\rmd^2\tau}{4\pi^2}\;
  \frac{1}{[1 + |\tau|^2\, (t_2/\beta_1 + (1-t_2)/\alpha_1)/4]^{N-1}}
\nonumber\\
= \frac{\mcP_{NK}(t_2)}{N-1}\;
  \frac{1}{\alpha_1^{N-K}\, \beta_1^K}\; \frac{1}{\pi (N-2)}\;
 \frac{1}{t_2/\beta_1 + (1-t_2)/\alpha_1} 
\nonumber\\
= \frac{\mcP_{NK}(t_2)}{\pi\, (N-1)(N-2)}\; 
   \frac{1}{\alpha_1^{N-K-1}\, \beta_1^{K-1}}\;
   \frac{1}{t_2\, \alpha_1 + (1-t_2)\, \beta_1} \; ,
\end{eqnarray}
where we have used the integration formula (\ref{app:LorentzInt2}) for complex 
Lorentzian integrals. From this it follows 
\begin{equation}
\fl \mcP_{NK}(t_1,t_2)= 
   \frac{\mcP_{NK}(t_2)}{\pi\, (N-1)(N-2)}
   \frac{(N-2)^2}{Z(\bi{0})}\int\frac{\rmd s_1}{2\pi}\;
   \frac{1}{\alpha_1^{N-K-1}\, \beta_1^{K-1}\, 
            [1 +\rmi s_1 (t_1+t_2 -1)]} \; .
\end{equation}
With
\begin{equation}
Z(\bi{0})= \int\frac{\rmd^2\tau}{4\pi^2} \frac{1}{(1+|\tau|^2/4)^{N}} 
 = \frac{1}{\pi\, (N-1)} \; ,
\end{equation}
we finally obtain
\begin{equation}
\mcP_{NK}(t_1,t_2)= \mcP_{NK}(t_2)\; \frac{N-2}{2\pi}\int
   \frac{\rmd s_1}{\alpha_1^{N-K-1}\, \beta_1^{K-1}\, 
            [1 +\rmi s_1 (t_1+t_2 -1)]} \; .
\label{V2U:finres}\end{equation}

\subsection{\label{V2A} Examples and applications}

\subsubsection*{Small dimensions}

\begin{figure}
\includegraphics[width=0.7\textwidth]{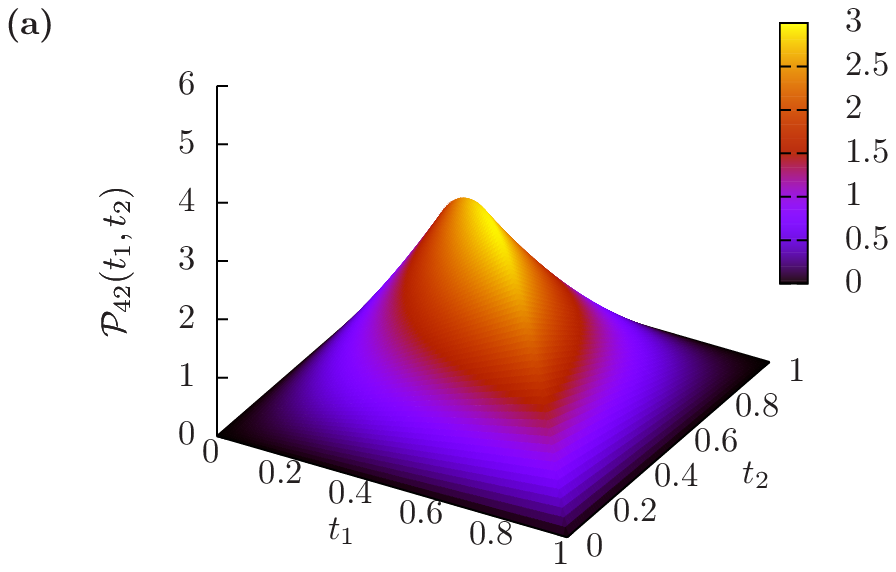}
\hspace{-35mm}\includegraphics[width=0.7\textwidth]{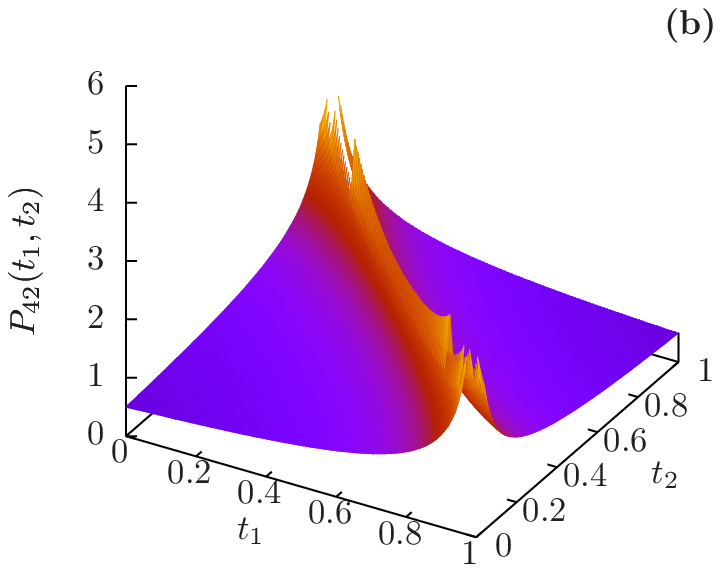}
\caption{Joint probability distributions for $N=4$ and $R=K=2$. Panel (a)
shows the unitary case, (\ref{V2A:mcP42res}); panel (b) shows the 
orthogonal case, (\ref{V2A:P42res}). The color box shown in panel (a) is
also valid for panel (b).}
\label{f:smalldims}\end{figure}

Here, we choose the dimensions as $N=4$ and $R=K=2$. The unitary case is much 
simpler than the orthogonal one, as the evaluation of the remaining integral
can be done by a straight forward application of the residue theorem. Namely,
from (\ref{V2U:finres}) we find
\begin{equation}
\mcP_{42}(t_1,t_2)= \frac{\mcP_{42}(t_2)}{\pi}\int\frac{\rmd s}
  {(1+\rmi t_1\, s)\, (1-\rmi (1-t_1)\, s)\, (1+\rmi (t_1+t_2-1)\, s)} \; .
\end{equation}
The integrand has three simple poles on the imaginary line, one pole above 
the point $\rmi$, the other pole below the point $-\rmi$, and the third pole
below $-\rmi$ (above $\rmi$) for $t_1+t_2 < 1$ ($t_1+t_2 > 1$). With
$\mcP_{42}(t_2)= 6\, t_2\, (1-t_2)$ obtained from (\ref{V1:Uresult2}), we find
\begin{equation}
\mcP_{42}(t_1,t_2)= 12\left\{ \begin{array}{ccr}
   t_1\, t_2 &:& t_1 + t_2 < 1\\
   (1-t_1)\, (1-t_2) &:& t_1 + t_2 > 1\end{array}\right. \; .
\label{V2A:mcP42res}\end{equation}
In the orthogonal case, the calculation is more involved. According to
(\ref{V2O:finres}) we have
\begin{equation}
P_{42}(t_1,t_2) = \frac{P_{42}(t_2)}{4\pi}\int\frac{\rmd s}{
   \sqrt{(1+\rmi t_1\, s)\, (1-\rmi (1-t_1)\, s)\, (1+\rmi (t_1+t_2-1)\, s)}}
 \; .
\end{equation}
Via the variable substitution $s\to \phi$ with 
$\tan\phi = 2\, [ t_1 (1-t_1)\, s+\rmi\, (t_1-1/2)]$, we arrive at
\begin{equation}
P_{42}(t_1,t_2)= \frac{1}{4\pi}\int_{-\pi/2}^{\pi/2}\frac{\rmd\phi}{
   \sqrt{(t_1 t_2 -t_{\rm s})\, \cos^2\phi + \rmi\, t_{\rm s}\, \sin\phi\,
      \cos\phi}}\; ,
\end{equation}
where $t_{\rm s}= (t_1+t_2-1)/2$ and where we have used that $P_{42}(t)= 1$,
cf. (\ref{V1:Oresult2}). Standard manipulations of the trigonometric
expressions then lead to
\begin{equation}
\fl P_{42}(t_1,t_2)= \frac{1}{\sqrt{2\, \sqrt{t_1 t_2 (1-t_1)(1-t_2)}\; 
   (1+\cos\alpha)}}\; \frac{1}{\pi}\int_0^{\pi/2}\frac{\rmd\phi}{
   \sqrt{1 -2(1+\cos\alpha)^{-1}\; \sin^2\phi}} \; ,
\end{equation}
where
$2\cos\alpha = [t_1 t_2 + (1-t_1)(1-t_2)]/\sqrt{t_1 t_2 (1-t_1)(1-t_2)}$. The
remaining integral is identical to the complete elliptic integral of the first
kind~\cite{AbrSte70}. This allows us to write $P_{42}(t_1,t_2)$ in the 
following form
\begin{equation}
P_{42}(t_1,t_2) = \frac{1}{\pi (a+b)}\; 
   K\Big [\, \frac{2\, \sqrt{ab}}{a+b}\, \Big ] \; ,
\label{V2A:P42res}\end{equation}
where $a= \sqrt{t_1 t_2}$ and $b= \sqrt{(1-t_1)(1-t_2)}$.

Figure~\ref{f:smalldims} shows the joint probability distributions for $N=4$ 
and $K=2$, for the unitary case in panel (a), and the orthogonal case in panel 
(b). The distribution function looks rather unspectacular in the unitary case.
However, the orthogonal case shows some peculiarities worthwhile mentioning:
First, there is a square root singularity in the $(t_1,t_2)$ plane as one 
approaches the line $t_1+t_2 = 1$, and second, the distribution function 
approaches finite values on the border of the $(t_1,t_2)$ unit square. This
is surprising, because anywhere outside the unit square, the distribution
function must be equal to zero, of course. We have verified both analytic
results with the help of random matrix simulations~\cite{Lazaro2015}. 

\subsubsection*{Asymmetric cases}

\begin{figure}
\includegraphics[width=0.7\textwidth]{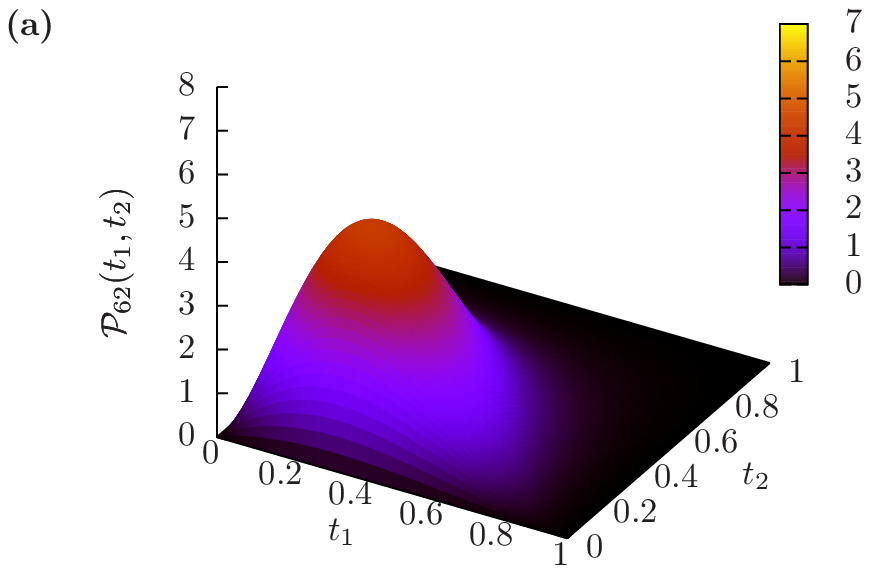}
\hspace{-35mm}\includegraphics[width=0.7\textwidth]{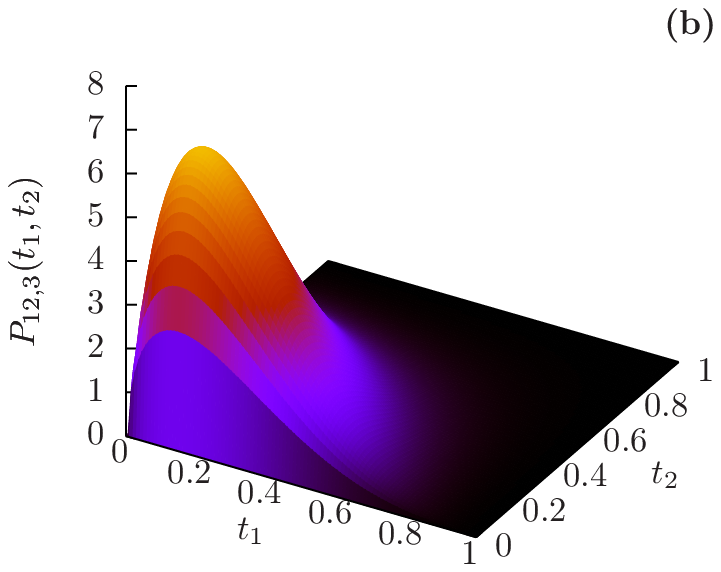}
\caption{Panel (a) shows the joint probability density for $N=6$ and 
$K=2$ for the unitary case, (\ref{V2A:mcP62res}); panel (b) shows the 
joint probability density for $N=12$, $K=3$ for the orthogonal case, 
(\ref{V2A:P12_3res}). The color box shown in panel (a) is
also valid for panel (b).}
\label{f:largedims1}\end{figure}

Let us start again with the unitary case. According to (\ref{V2U:finres})
$\mcP_{NK}(t_1,t_2)= \mcP_{NK}(t_2)\; I_{N-K-1,K-1}(t_1,t_2)$, with
\begin{equation}
I_{mk}(t_1,t_2)= \frac{m+k}{2\pi}\int\frac{\rmd s}{(1+\rmi s t_1)^m\,
   (1-\rmi s (1-t_1))^k\, (1+\rmi s t_{\rm s})} \; ,
\end{equation}
being the integral left to be evaluated, and $t_{\rm s}= t_1+t_2-1$. Still, the
integrand has three poles on the imaginary axis, at $\rmi/t_1$ (of order $m$), 
another at $-\rmi/(1-t_1)$ (of order $k$), and a third simple pole at 
$\rmi/t_{\rm s}$. Applying the residue theorem, we find the following: For 
$t_{\rm s} < 0$, the pole at $\rmi/t_1$ is the only pole in the upper half 
plane, such that
\begin{equation}
I_{mk}(t_1,t_2)= \rmi\, (m+k)\; {\rm Res}\big [\, \alpha_1^{-m}\, 
   \beta_1^{-k}\, (1+\rmi s t_{\rm s})^{-1}, \rmi/t_1\, \big ] \; .
\end{equation}
For $t_{\rm s} > 0$, the pole at $-\rmi/(1-t_1)$ is the only pole in the 
lower half plane, such that 
\begin{equation}
I_{mk}(t_1,t_2)= -\rmi\, (m+k)\; {\rm Res}\big [\, \alpha_1^{-m}\, 
   \beta_1^{-k}\, (1+\rmi s t_{\rm s})^{-1}, -\rmi/(1-t_1)\, \big ] \; .
\end{equation}
We calculate these residues with the help of a computer algebra 
system~\cite{Maxima2014}. This allows to obtain exact analytic expressions even 
for large integers $N$ and $K$. Here, we choose $N=6$ and $K=2$ and obtain
\begin{equation}
\mcP_{62}(t_1,t_2)= 80\left\{ \begin{array}{ccr}
   t_1 t_2\, S_{31}(t_1,t_2) &:& t_1+t_2 < 1\\
   (1-t_1)^3\, (1-t_2)^3 &:& t_1+t_2 > 1\end{array}\right. \; ,
\label{V2A:mcP62res}\end{equation}
where $S_{31}(t_1,t_2)= t_1 t_2 (t_1 t_2 + 9) + 3\, t_1^2\, (1-t_2) 
  + 3\, t_2^2\, (1-t_1) -6\, (t_1+t_2) + 3$. This probability density is shown 
in figure~\ref{f:largedims1}, panel (a). 

Now, let us consider the orthogonal case. Starting from (\ref{V2O:finres}),
we apply the substitution $s_1 \to z= \sqrt{1 + \rmi s_1 (t_1+t_2-1)}$ to
obtain
\begin{equation}
 P_{NK}(t_1,t_2)= P_{NK}(t_2)\; J_{(N-K-1)/2,(K-1)/2}(t_1,t_2)\; ,
\end{equation}
where
\begin{equation}
\fl J_{mk}(t_1,t_2)= \frac{m+k-1/2}{\rmi\pi}\;
   \frac{t_{\rm s}^{m+k-1}}{t_1^m (1-t_1)^k} \int\frac{\rmd z}
   {[z^2 - (1-t_2)/t_1]^m\, [ t_2/(1-t_1) -z^2]^k} \; .
\end{equation}
Here, we can again apply the residue theorem, but only when $m$ and $k$ are
both integers. This means that $K$ must be odd and $N$ must be even. Below, we
will assume that this is the case.

For $t_{\rm s} = t_1+t_2-1 > 0$, the integration path comes from infinity, 
following the diagonal in the lower right quadrant of the complex plane, 
approaching the origin it leaves the diagonal upwards to cross the real axis at 
$z=1$. Then the path continues towards infinity again, approaching 
the diagonal in the upper right quadrant. For $m+k \ge 1$, this integration 
path can be considered as a closed loop without changing the value of the
integral. Moreover, since $t_{\rm s} > 0$ implies 
$t_2/(1-t_1) > 1 > (1-t_2)/t_1$, the only pole inside this loop is at  
$b= \sqrt{t_2/(1-t_1)}$. Therefore, we find with $a= \sqrt{(1-t_2)/t_1}$:
\begin{equation}
\fl J_{mk}^+(t_1,t_2)= (m+k-1/2)\;
   \frac{-2\; t_{\rm s}^{m+k-1}}{t_1^m (1-t_1)^k}\;
   {\rm Res}[ (z^2 - a^2)^{-m}\, (b^2 - z^2)^{-k}, b] \; ,
\end{equation}
where the minus sign comes from the fact that the orientation of the path is
mathematically negative. For $t_{\rm s} < 0$ the orientation of the integration 
path changes sign. Furthermore, $t_{\rm s} < 0$ implies $b < 1 < a$ such that 
the only pole within the integration path is now at $a$. Therefore, we find for 
this case
\begin{equation}
\fl J_{mk}^-(t_1,t_2)= (m+k-1/2)\;
   \frac{2\; t_{\rm s}^{m+k-1}}{t_1^m (1-t_1)^k}\;
   {\rm Res}[ (z^2 - a^2)^{-m}\, (b^2 - z^2)^{-k}, a] \; .
\end{equation}
As an example, let us choose $N=12$ and $K=3$. Using \cite{Maxima2014} again, 
we find
\begin{equation}
P_{12,3}(t_1,t_2)= \frac{72}{7\pi}\left\{ \begin{array}{ccr}
   \sqrt{t_1 t_2}\; \mathcal{S}_{12,3} &:& t_1+t_2 < 1\\
   16\; (1-t_1)^{7/2}\, (1-t_2)^{7/2} &:& t_1+t_2 > 1
  \end{array}\right. \; ,
\label{V2A:P12_3res}\end{equation}
where $\mathcal{S}_{12,3}= 16\, t_1^3 t_2^3 - 56\, t_1^2 t_2^3
   + 70\, t_1 t_2^3 - 35\, t_2^3 - 56\, t_1^3 t_2^2 + 196\, t_1^2 t_2^2
   - 245\, t_1 t_2^2 + 105\, t_2^2 + 70\, t_1^3 t_2 - 245\, t_1^2 t_2 
   + 280\, t_1 t_2 - 105\, t_2 - 35\, t_1^3 + 105\, t_1^2 - 105\, t_1 + 35$.
This probability density is shown in figure~\ref{f:largedims1}, panel (b). 

For the orthogonal case, we choose values for $N$ and $K$ which are 
approximately twice as large as in the unitary case. The reason is that for
the one-point functions it holds that $\mcP_{NK}(t) = P_{2N,2K}(t)$ as 
discussed before (\ref{V1:Uresult}). Thus one could have expected that the 
probability densities $\mcP_{62}(t_1,t_2)$ and $P_{12,3}(t_1,t_2)$ would at 
least look similar. Note though that we choose $P_{12,3}(t_1,t_2)$ instead of 
$P_{12,4}(t_1,t_2)$ because in this case it is much easier to evaluate the 
integral in (\ref{V2O:finres}), as explained above. Comparing the two cases in
figure~\ref{f:largedims1}, one can clearly see that $P_{12,3}(t_1,t_2)$, shown 
in panel (b), has much steeper slopes as $t_1$ or $t_2$ tend to zero, than 
$\mcP_{62}(t_1,t_2)$, shown in panel (a).

\subsubsection*{Measurement outcomes for orthonormal states}

At last, we would like to mention two instances, where our results can describe
statistical properties of real physical systems. The first example, discussed
here, is about the probabilities of a measurement outcome for different 
eigenstates of a quantum system, for which the quantum chaos conjecture 
applies~\cite{BerTab77,CVG80,BGS84,BluSmi88}. When applied to the eigenstates
of a Hamiltonian quantum system, the conjecture states that in the 
semiclassical limit these eigenstates will have the same statistical properties 
as the column vectors of elements from $O(N)$ or $U(N)$, depending on whether 
the system has an anti-unitary symmetry (usually related to time-reversal) or 
not~\cite{Haake2001,Stoe99}. We will call the Hamiltonian of such a quantum 
system a "quantum-chaotic" Hamiltonian.

In this situation, we consider the $K$-fold degenerate measurement outcome of
a projective measurement, as described in (\ref{I:normsquaredproj}) 
and (\ref{I:txiprojs}). Then, our results describe the joint probability 
distribution for the probabilities $t_\xi$. In order to compare experimental
data or numerical simulations with our results, one needs to perform an average
over samples of different eigenstates and different systems. Samples of 
different quantum-chaotic systems may be obtained by changing some parameter of 
the Hamiltonian taking care to remain in a region where the quantum chaos 
conjecture still applies.

Our results can be used also, if the system is prepared in a mixture of 
eigenstates of the quantum chaotic Hamiltonian. That might be a thermal state 
with canonical distribution. Such a mixture may be described by the density
matrix
\begin{equation}
\varrho = \sum_{\xi=1}^R p_\xi \; |\psi_\xi\ra\la\psi_\xi| \; , \qquad
\sum_{\xi=1}^R p_\xi = 1 \; .
\end{equation}
Then, the probability of the above measurement outcome is given by
\begin{equation}
\bar t = {\rm tr}\big ( \hat P_1\, \varrho\, \big )
 = \sum_{\xi=1}^R p_\xi\; t_\xi \; ,
\end{equation}
where $\hat P_1$ denotes the projector on the $K$-dimensional eigenspace, 
corresponding to that measurement outcome. In order to calculate the 
probability distribution of $\bar t$, for fixed but arbitrary occupation 
probabilities $p_\xi$, the joint probability density for the individual $t_\xi$
would be the ideal starting point. 

The quantum chaos conjecture has been extended to classical wave systems, also. 
Examples are two-dimensional microwave cavities~\cite{StoSte90} and 
one-dimensional microwave networks~\cite{Hul04}, elastomechanic 
systems~\cite{Elleg95}, and acoustic waves~\cite{Der95}, among others. In those 
cases, the main difference is that the $t_\xi$ will describe intensities rather 
than probabilities.

\subsubsection*{Universal conductance fluctuations}

In this area, one is interested in the distribution of the conductance of
charge carriers through mesoscopic structures~\cite{LeeSto85,Altsh85}. The 
theoretical description is based on the Landauer-B\" uttiker 
formula~\cite{BueLan85}, and a maximum entropy principle~\cite{MelSel80}, which 
predicts the transport properties to be statistical in nature, and described by 
appropriate ensembles of scattering matrices~\cite{Bee97,MelKum04}. At present, 
we can make contact with the statistics of conductances in the unitary case, 
only. Again, this is the case which describes systems without time-reversal 
symmetry.

Hence, consider a two terminal scattering problem, with $K$ modes in
one lead and $N-K$ modes in the other lead. Then, according to the 
Landauer-B\" uttiker formalism, the scattering matrix $\bi{S}$ is a 
$N$$\times$$N$-matrix where the off-diagonal $K$$\times$$(N-K)$ dimensional
block-matrix $\btau$ contains the amplitudes for transitions from the modes
of one lead to those of the other lead. In terms of the elementary conductance
unit $g_0= 2e^2/h$, the conductance is given by the simple formula 
$g= {\rm tr}(\btau^\dagger\, \btau)$. In the universal regime, i.e. when 
the maximum entropy principle applies, $\bi{S}$ may be taken as a random matrix 
from $U(N)$ provided with the Haar measure, which is the reason that our 
results apply. Usually, the statistical properties of the conductance are 
computed from the distribution of the eigenvalues of 
$\btau^\dagger\, \btau$. However, it is also possible to express $g$ as a
sum of partial conductances
\begin{equation}
g= \sum_{\xi=1}^{N-K} \sum_{j=1}^K |\tau_{j\xi}|^2 
 = \sum_{\xi=1}^{N-K} t_\xi \; .
\end{equation}
In distinction to the eigenvalues of $\btau^\dagger\, \btau$, the partial 
conductances $t_\xi$ may be directly measurable by mode-selective measurements.
Also, just as in the case of the closed systems discussed above, there might be
situations where the total conductance is given by a weighted sum of the 
partial conductances for instance when the modes in one lead are occupied with
different probabilities according to some temperature profile. In such a case,
$g$ can no longer be expressed in terms of the eigenvalues of 
$\btau^\dagger\, \btau$.

In~\cite{KumPan10}, the authors collected known results and presented a 
new approach to calculate the distribution of the conductance in two-terminal 
quantum transport with an arbitrary number of modes in each lead. As an 
illustration, we consider the case of a scattering system with two modes in 
each lead, which corresponds to our unitary case with $N=4$ and $K=2$, 
considered above. In this case, the quantities $t_1$ and $t_2$ are the 
partial conductances such that $g= t_1+t_2$. Then, starting from the 
probability density given in (\ref{V2A:mcP42res}) we can recover $p(g)$ 
from~\cite{KumPan10} as
\begin{eqnarray}
p(g) &= \int\rmd t_1\rmd t_2\; \delta(g-t_1-t_2)\; \mcP_{42}(t_1,t_2)
    = \int_0^1\rmd t\; \mcP_{42}(t,g-t) \nonumber\\
    &= \left\{ \begin{array}{ccr}
         2\, g^3 &:& 0 \le g \le 1\\
         2\, (2-g)^3 &:& 1\le g\le 2 \end{array}\right. \; .
\end{eqnarray}

\section{\label{C} Conclusions}

In this paper, we considered the partial sums of absolute values squared of
a random orthonormal basis with $R$ elements in a $N$ dimensional vector
space. We derived general expressions for the joint probability density of 
these partial sums, and explained how these results can be related to 
experiments. Distinguishing between the vectors being real (orthogonal case) or
complex (unitary case), the general results are given in (\ref{VRO:genform}) 
and (\ref{VRU:genform}), respectively. They still involve $R (R+1)/2$ 
integrals, but for $R=2$, we could eventually evaluate all integrals and 
arrive at explicit analytic expressions. 

Obviously, the joint probability distributions are important only as long as
correlations between the partial sums are important. Otherwise the one-point
functions would be enough to describe their statistical properties. For small
dimensions, as the ones considered in our examples, such correlations are 
present. However, for increasing $N$ and $K$, it is natural to expect that 
correlations become less important. 

\ack
We acknowledge financial support from grant CONACyT CB-2009/129309.

\appendix

\section*{Appendix: Integration formulas}
\setcounter{section}{1}

\subsection*{Integral related to the one vector case}

\begin{equation}
\fl I_{mk}(a,b)= (m+k-1)\int\frac{\rmd s}{2\pi}\; 
   \frac{1}{(1+\rmi a\, s)^m\, (1-\rmi b\, s)^k} 
 = \frac{\Gamma(m+k)}{\Gamma(m)\, \Gamma(k)}\; 
   \frac{a^{k-1}\, b^{m-1}}{(a+b)^{m+k-1}} \; .
\label{app:Imk}\end{equation}
This integral is well defined for $m,k$ being integers or half integers, with
$m+k \ge 2$ and $0 < a,b < 1$. To prove this integration formula, we first use
integration by parts to demonstrate
\begin{equation}
I_{mk}(a,b)= \frac{a\, m}{b\, (k-1)}\; I_{m+1,k-1}(a,b) \; .
\end{equation}
This yields, in the case of integer $k$:
\begin{equation}
I_{mk}(a,b)= \frac{\Gamma(m+k-1)}{\Gamma(m)\, \Gamma(k)}\; 
   \Big (\frac{a}{b}\Big )^{k-1}\; I_{n-1,1}(a,b) \; .
\label{app:ImkInt}\end{equation}
For $k$ being a half integer ($2k$: odd) we find instead
\begin{equation}
I_{mk}(a,b)= \frac{\Gamma(m+k-1/2)}{\Gamma(m)\, \Gamma(k)}\; 
   \Big (\frac{a}{b}\Big )^{k-1/2}\; I_{n-1/2,1/2}(a,b) \; .
\label{app:ImkHalfInt}\end{equation}
In the first case, (\ref{app:ImkInt}), we directly use the residue
theorem to evaluate $I_{n-1,1}(a,b)$, while in the 
second case, (\ref{app:ImkHalfInt}), we first apply the variable 
transformation $s \to z=\sqrt{1-\rmi\, bs}$, before applying the residue 
theorem. In both cases, the results lead to the same formula as given in 
(\ref{app:Imk}). For the orthogonal and the unitary case, we find 
respectively
\begin{equation}
P_{NK}(t)= I_{(N-K)/2,K/2}(t,1-t) \; , \qquad 
\mcP_{NK}(t)= I_{N-K,K}(t,t-1)\; .
\label{app:PNKImk}\end{equation}

\subsection*{A generalization of the integral above}

For the case $R=2$, we need a generalized version of the integration formula in
(\ref{app:Imk}). This is
\begin{equation}
I_{mk}(\alpha,a;\beta,b)= (m+k-1)\int\frac{\rmd s}{2\pi}\;
   \frac{1}{(\alpha + \rmi a\, s)^m\, (\beta -\rmi b\, s)^k} \; ,
\end{equation}
for complex parameters $\alpha,\beta$ with real parts larger than one. With the
help of simple algebraic manipulations it can be shown that
\begin{equation}
I_{mk}(\alpha,a;\beta,b)= I_{mk}(a,b)\; 
   \left( \frac{a+b}{\beta\, a + \alpha\, b}\right)^{m+k-1} \; .
\label{app:Iabmk}\end{equation}

\subsection*{Lorentzian integrals}

With $m \ge 1$, we obtain
\begin{equation}
\frac{1}{2\pi}\int\frac{\rmd\tau}{(1+c\, \tau^2)^m} 
  = \frac{\Gamma(m-1/2)}{2\sqrt{\pi\, c}\; \Gamma(m)} \; .
\label{app:LorentzInt}\end{equation}
This formula can be obtained from the residue theorem. For the unitary case, we 
also need the following two-dimensional (complex) version:
\begin{eqnarray}
\fl \frac{1}{4\pi^2}\int\frac{\rmd^2\tau}{(1+c\, |\tau|^2)^m} 
  = \frac{1}{4\pi^2}\int_0^{2\pi}\rmd\phi\int_0^\infty\rmd r\;
    \frac{r}{(1+c\, r^2)^m}
  = \frac{1}{2\pi}\; \frac{(1+c r^2)^{1-m}}{2c\, (1-m)}\Big |_{r=0}^\infty
\nonumber\\
 = \frac{1}{4\pi\, c\, (m-1)} \; .
\label{app:LorentzInt2}\end{eqnarray}

\subsection*{Delta function for complex arguments}

For $w\in\mathbb{C}$, we mat write
\begin{eqnarray}
\fl \delta^2(w)= \delta({\rm Re}(w)\, )\; \delta({\rm Im}(w)\, ) 
  = \int\frac{\rmd x\, \rmd y}{4\pi^2}\; 
    \rme^{-\rmi [y\, {\rm Re}(w) + x\, {\rm Im}(w)]}
  = \int\frac{\rmd x\, \rmd y}{4\pi^2}\; 
    \rme^{-\rmi {\rm Im}[(x+\rmi y)\, w]} \nonumber\\
  = \int\frac{\rmd^2 z}{4\pi^2}\; \rme^{-\rmi\, {\rm Im}(zw)} \; .
\label{app:2Dcomplexdelta}\end{eqnarray}

\section*{References}

\bibliographystyle{iopart-num}
\bibliography{/home/gorin/Bib/JabRef}

\providecommand{\newblock}{}
\begin{thebibliography}{10}
\expandafter\ifx\csname url\endcsname\relax
  \def\url#1{{\tt #1}}\fi
\expandafter\ifx\csname urlprefix\endcsname\relax\def\urlprefix{URL }\fi
\providecommand{\eprint}[2][]{\url{#2}}

\bibitem{Weyl39}
Weyl H 1939 {\em The classical groups\/} (Princton: Princton University Press)

\bibitem{Haar33}
Haar A 1933 {\em Ann. Math.\/} {\bf 34} 147

\bibitem{Meh2004}
Mehta M~L 2004 {\em Random matrices and the statistical theory of energy
  levels, 3rd Edition\/} (New York: Academic Press)

\bibitem{Bro81}
Brody T~A, Flores J, French J~B, Mello P~A, Pandey A and Wong S~S~M 1981 {\em
  Rev. Mod. Phys.\/} {\bf 53} 385--479

\bibitem{Haake2001}
Haake F 2001 {\em Quantum signatures of chaos, 2nd. Edition\/} (Berlin: Springer)

\bibitem{Stoe99}
St\" ockmann H~J 1999 {\em Quantum chaos\/} (Cambridge: Cambridge University
  Press)

\bibitem{Bee97}
Beenakker C~W~J 1997 {\em Rev. Mod. Phys.\/} {\bf 69} 731--808

\bibitem{MelKum04}
Mello P~A and Kumar N 2004 {\em Quantum transport in mesoscopic systems:
  Complexity and statistical fluctuations: A maximum-entropy viewpoint\/}
  (Oxford: Oxford University Press)

\bibitem{GS02b}
Gorin T and Seligman T~H 2002 {\em J. Opt. B: Quantum Semiclass. Opt.\/} {\bf
  4} S386--S392 

\bibitem{GS03}
Gorin T and Seligman T~H 2003 {\em Phys. Lett. A\/} {\bf 309} 61--67

\bibitem{PinSel15}
Pineda C and Seligman T~H 2015 {\em J. Phys. A: Math. Theor.\/} {\bf 48} 425005

\bibitem{Min05}
Mintert F, Carvalho A~R~R, K{\' u}s M and Buchleitner A 2005 {\em Phys. Rep.\/}
  {\bf 415} 207--260

\bibitem{Hor09}
Horodecki R, Horodecki P, Horodecki M and Horodecki K 2009 {\em Rev. Mod.
  Phys.\/} {\bf 81} 865--942

\bibitem{Aub03}
Aubert S and Lam C~S 2003 {\em J. Math. Phys.\/} {\bf 44} 6112--6131

\bibitem{ColSni06}
Collins B and {\' S}niady P 2006 {\em Commun. Math. Phys.\/} {\bf 264} 773--795

\bibitem{GL08}
Gorin T and Lopez G~V 2008 {\em J. Math. Phys.\/} {\bf 49} 013503 

\bibitem{ZycSom00}
Zyczkowski K and Sommers H~J 2000 {\em J. Phys. A: Math. Gen.\/} {\bf 33} 2045--2067

\bibitem{Nov07}
Novak J 2007 {\em The electronic journal of combinatorics\/} {\bf 14} R21

\bibitem{MelSel80}
Mello P~A and Seligman T~H 1980 {\em Nucl. Phys. A\/} {\bf 344} 489--508

\bibitem{MePeKu88}
Mello P, Pereyra P and Kumar N 1988 {\em Annals of Physics\/} {\bf 181} 290--317 

\bibitem{Kum13}
Kumar S, Nock A, Sommers H~J, Guhr T, Dietz B, Miski-Oglu M, Richter A and
  Sch\" afer F 2013 {\em Phys. Rev. Lett.\/} {\bf 111} 030403

\bibitem{NoKSG14}
Nock A, Kumar S, Sommers H~J and Guhr T 2014 {\em Annals of Physics\/} 
  {\bf 342} 103--132 

\bibitem{UllPor63}
Ullah N and Porter C~E 1963 {\em Phys. Rev.\/} {\bf 132} 948--950

\bibitem{Ullah64}
Ullah N 1964 {\em Nucl. Phys.\/} {\bf 58} 65--71

\bibitem{Gor02}
Gorin T 2002 {\em J. Math. Phys.\/} {\bf 43} 3342--3351

\bibitem{AbrSte70}
Abramowitz M and Stegun I~A (eds) 1970 {\em Handbook of mathematical
  functions\/} (New York: Dover publications, inc.)

\bibitem{Lazaro2015}
Alonso L (unpublished) {\em 
  Joined probability distributions for quantum transport\/} Ph.D. thesis
  Universidad de Guadalajara

\bibitem{Maxima2014}
 2014 Maxima, a computer algebra system. version 5.35.1
  \urlprefix\url{http://maxima.sourceforge.net}

\bibitem{BerTab77}
Berry M~V and Tabor M 1977 {\em Proc. R. Soc. Lond. A\/} {\bf 356} 375--394

\bibitem{CVG80}
Casati G, Valz-Gris F and Guarneri I 1980 {\em Lett. Nuovo Cimento\/} {\bf 28}
  279

\bibitem{BGS84}
Bohigas O, Giannoni M~J and Schmit C 1984 {\em Phys. Rev. Lett.\/} {\bf 52}
  1--4

\bibitem{BluSmi88}
Bl{\" u}mel R and Smilansky U 1988 {\em Phys. Rev. Lett.\/} {\bf 60} 477--480

\bibitem{StoSte90}
St\"ockmann H~J and Stein J 1990 {\em Phys. Rev. Lett.\/} {\bf 64}
  2215--2218

\bibitem{Hul04}
Hul O, Bauch S, Pako\ifmmode~\acute{n}\else \'{n}\fi{}ski P, Savytskyy N,
  \ifmmode~\dot{Z}\else \.{Z}\fi{}yczkowski K and Sirko L 2004 {\em Phys. Rev.
  E\/} {\bf 69} 056205

\bibitem{Elleg95}
Ellegaard C, Guhr T, Lindemann K, Lorensen H~Q, Nyg\aa{}rd J and Oxborrow M
  1995 {\em Phys. Rev. Lett.\/} {\bf 75} 1546--1549

\bibitem{Der95}
Derode A, Roux P and Fink M 1995 {\em Physical Review Letters\/} {\bf 75}
  4206--4209

\bibitem{LeeSto85}
Lee P~A and Stone A~D 1985 {\em Phys. Rev. Lett.\/} {\bf 55} 1622--1625

\bibitem{Altsh85}
Altshuler B~L 1985 {\em Pis'ma Zh. Eksp. Teor. Fiz.\/} {\bf 41} 530 [JETP Lett.
  41: 648]

\bibitem{BueLan85}
B\"uttiker M, Imry Y, Landauer R and Pinhas S 1985 {\em Phys. Rev. B\/} {\bf
  31}(10) 6207--6215

\bibitem{KumPan10}
Kumar S and Pandey A 2010 {\em J. Phys. A: Math.  Theor.\/} {\bf 43} 285101

\end{thebibliography}
 
\end{document}